\documentclass[superscriptaddress, amsmath,amssymb, pre, twocolumn,floatfix]{revtex4-1}

\usepackage{graphicx}
\usepackage{dcolumn}
\usepackage{bm}
\usepackage{hyperref}
\usepackage{color}
\usepackage{amsmath}
\usepackage{amssymb}

\usepackage{mathtools}
\usepackage{xfrac}

\usepackage{bibentry}
\usepackage{endnotes}

\begin{document}

\title{Stiffening Thermal Membranes by Cutting}
\author{Emily Russell}
\affiliation{ICERM, Box 1955, Brown University, Providence, RI 02912, USA}
\email{Emily\_Russell@post.harvard.edu}
\author{Rastko Sknepnek}
\affiliation{School of Science and Engineering, University of Dundee, Dundee DD1 4HN, United Kingdom} 
\affiliation{School of Life Sciences, University of Dundee, Dundee DD1 4HN, United Kingdom} 
\author{Mark Bowick}
\affiliation{Department of Physics and Soft Matter Program, Syracuse University, Syracuse, NY 13244, USA}
\affiliation{Kavli Institute for Theoretical Physics, University of California, Santa Barbara, CA 93106, USA}
\email{mjbowick@syr.edu}

\date{\today}

\begin{abstract}

Two-dimensional crystalline membranes have recently been realized experimentally in such systems as graphene and molybdenum disulfide, 
sparking a resurgence in interest in their statistical properties.  Thermal fluctuations can significantly affect the effective mechanical properties
of properly thermalized membranes, renormalizing both bending rigidity and elastic moduli so that in particular they become stiffer to bending than their bare bending rigidity would suggest.  We use molecular dynamics simulations to examine how the mechanical behavior of thermalized two-dimensional clamped ribbons (cantilevers) depends on their precise topology and geometry. We find that a simple slit smooths roughness as measured by the variance of height fluctuations. This counterintuitive effect may be due to the counter-posed coupling of the lips of the slit to twist in the intact regions of the ribbon.

\end{abstract}

\maketitle

\section{Introduction}

Two-dimensional materials have recently garnered interest not only for their remarkable electronic properties \cite{novoselov2005two}, but also for their mechanical behavior.  A wide variety of atomically thin materials  are now known, including graphene, molybdenum disulfide~\cite{wang2012electronics},
boron nitride~\cite{song2010large} and black phosphorus~\cite{li2014black}. All these materials have very unusual electronic and mechanical properties \cite{geim2013van}. Graphene, for example, is very strong yet flexible and resilient \cite{min2011mechanical,ovid2013mechanical}, making it possible to cut and fold sheets into complex three-dimensional structures \cite{Blees_Nature_2015}. It should be possible to design meta-materials with targeted properties for potential applications ranging from electronic devices to biological sensors. To develop the design principles required one must have a thorough understanding of the mechanical properties of such
materials and how they are affected by the detailed geometry and topology of  a sliced and punctured sheet. 

The mechanical properties of atomically thin materials are heavily influenced by thermal fluctuations, primarily because they much prefer to bend than to stretch. This was first examined in the context of thermalized elastic or crystalline  membranes by Nelson and Peliti and others in the late 1980s \cite{Nelson_JPhys_1987,aronovitz1988fluctuations}. Using low temperature perturbation theory Nelson and Peliti found that the nonlinear coupling of thermally excited out-of-plane deformations to in-plane phonons for  membranes with an in-plane shear modulus generates an effective long-range 2-point interaction between points with Gaussian curvature. The key result is an effective bending rigidity that is scale-dependent, growing almost linearly in the size over which the membrane is bent \cite{Nelson_JPhys_1987,nelson2004statistical, Bowick_PhysRep_2001, paulose2012fluctuating}.  More specifically the bending rigidity $\kappa$ scales as $\sim L^{\eta}$, where $L$ is a measure of the spatial extent of the shape fluctuations \cite{Nelson_JPhys_1987,Kantor_PRA_1987}. More elaborate  self-consistent field theory calculations that take account of the simultaneous scale dependence of the elastic moduli give $\eta\approx0.821$ \cite{le1992self}. Other analytical approaches lead to refined values of $\eta$~\cite{gazit2009structure,kownacki2009crumpling,braghin2010thermal,troster2013high,troster2015fourier} and allow comparisons with simulation results \cite{bowick1996flat,los2009scaling}. Possible evidence of such thermal renormalization of the bending rigidity has only recently been observed experimentally in graphene \cite{Blees_Nature_2015}.

Here we examine the consequences of making a simple straight cut down the middle of a thin thermalized semi-flexible ribbon. Slits allow for new configurations  that couple bending and stretching deformations \cite{Blees_Nature_2015,yllanes2017thermal}. The effective spring constant of a sheet with an array of slits is controlled by the number, size, and spacing of the slits \cite{Blees_Nature_2015}. We find counter-intuitive behavior even in this very simple setup. The root-mean-square height fluctuations are suppressed, effectively stiffening the ribbon to macroscopic bend with the more stiffening as the slit is lengthened. 

This paper is organized as follows. In Sec.~\ref{sec:model} we introduce a coarse grained, triangular lattice-based model for a two-dimensional elastic membrane. In Sec.~\ref{sec:results} we present
results of molecular dynamics simulations of ribbons with various slit lengths. Finally, in Sec.~\ref{sec:discussion} we discuss the phenomenology of the effects we observed and give the outlook of the 
open questions remaining.  

\section{Model}
\label{sec:model}

\begin{figure*}
\centering
\includegraphics[width=0.95\textwidth]{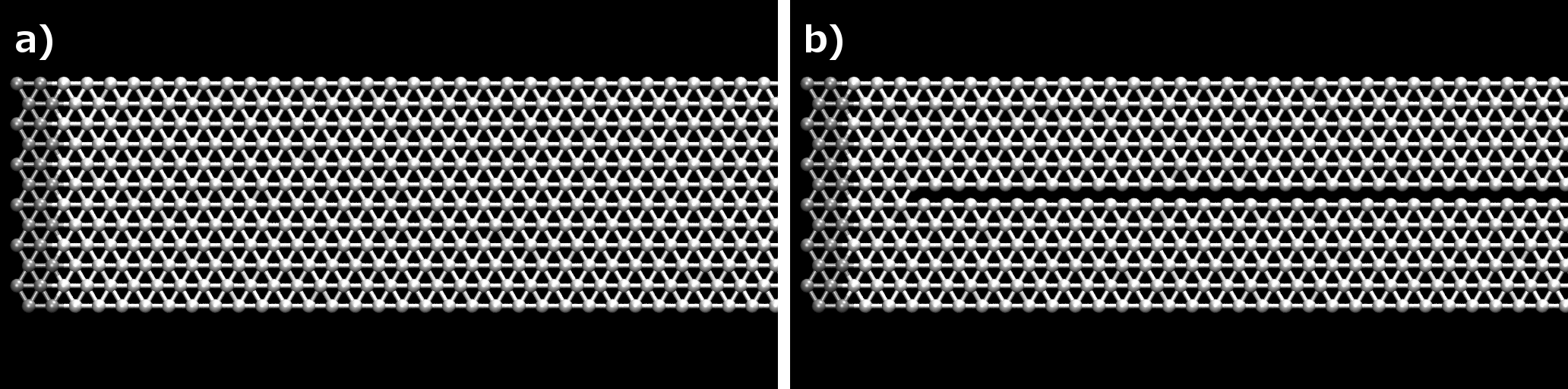}
\caption{Triangulation of a two-dimensional membrane for molecular dynamics simulations.  a) Portion of the triangulation of a simple rectangular ribbon of width $W=10a$.  
b) Modification of the triangulation in the presence of a slit. Dark gray vertices are held fixed during simulations, as described in the text.}
\label{Fig_triangulation}
\end{figure*}

We consider a generic two-dimensional crystalline membrane, with both a stretching energy and a bending energy, undergoing thermal fluctuations.  In order to carry out explicit molecular 
dynamics simulations, we discretize the membrane by triangulating it into equilateral triangles of side length $a \equiv 1.0$. Note that this is a dual lattice of the hexagonal lattice of 
graphene sheets, i.e., placing a carbon atom at the center of each triangle acts as a reliable approximate atomic model for the elastic modes of graphene \cite{bowick2017non}, as long as 
one chooses the bending rigidity and Young's modulus correctly. In the simulations reported here, we work primarily with rectangular ribbons of length $L\approx100a$ and width 
$W\approx10a$, yielding 1188 vertices of the triangulation (Fig.~\ref{Fig_triangulation}a).  These vertices are treated as particles of mass $m$ in a molecular dynamics simulation.  
We hold fixed the first two rows of vertices, clamping the ribbon at one end; this prevents overall motion of the ribbon during the simulation and imposes a boundary conditions on the 
normals to the ribbon surface, and thus reflects common experimental techniques \cite{Blees_Nature_2015}.

The stretching energy term \cite{Seung_PRA_1988}, representing the effect of the (two-dimensional) Young modulus $Y$ of the membrane, is given by introducing harmonic bonds 
between adjacent vertices:
\begin{equation}
E_{stretch} = \sum_{\langle i, j \rangle} \frac{1}{2} k (r_{ij}-a)^2\label{eq:stretching}
\end{equation}
with the sum taken over all pairs of adjacent vertices $i$ and $j$; $r_{ij}$ is the distance between the vertices and $k = 3600\sfrac{k_BT}{a^2}$.  The geometry of the triangulation relates this spring constant $k$ to the continuous Young modulus $Y$ by $Y = \frac{2}{\sqrt{3}}k$ \cite{Seung_PRA_1988}. Note that this value of the spring constant makes the system effectively unstretchable. 

The bending energy \cite{Seung_PRA_1988}, representing the effect of the bending rigidity $\kappa$ of the membrane, is given by introducing an energy cost to changing the dihedral 
angle between adjacent faces of the triangulation:
\begin{equation}
E_{bend} = \sum_{\langle \alpha, \beta \rangle} \tilde{\kappa} (1-\hat{n}_\alpha \cdot \hat{n}_\beta) = \sum_{\langle \alpha, \beta \rangle} \tilde{\kappa} (1 + \cos(\theta_{\alpha\beta})),\label{eq:bending}
\end{equation}
with the sum taken over all pairs of adjacent \emph{faces} $\alpha$ and $\beta$; $\hat{n}_\alpha$ and $\hat n_\beta$ are the normals to the faces; $\theta_{\alpha\beta}$ the dihedral angle between the faces; 
and $\tilde{\kappa}=5k_BT$ in most simulations. To explore the effects of varying bending rigidity, we have also performed several simulations for $\tilde{\kappa}$ in the range $5-80k_BT$.  
The coupling $\tilde{\kappa}$ is related to the continuous bending rigidity $\kappa$ by $\kappa = \frac{\sqrt{3}}{2} \tilde{\kappa}$ \cite{Seung_PRA_1988, Schmidt_JMPS_2012}. 
In a molecular dynamics simulation we need to compute the force on each particle. The expression for the bending force is lengthy but can be straightforwardly obtained by computing the 
gradient of Eq.~\eqref{eq:bending} with respect to the vertex position. Both stretching and bending energy given in the above form are often referred to as 
``harmonic'' and ``dihedral'' potential and are standard in most molecular dynamics packages. We note, however, that some software packages introduce a prefactor $\sfrac{1}{2}$ in the dihedral energy in Eq.~(\ref{eq:bending}).

These parameters give a \emph{F\"oppl von K\'arm\'an number} approximately in the range, 
\begin{equation}
\rm{vK} \equiv \frac{YLW}{\kappa} \approx 10^5 - 2 \times 10^6.
\end{equation}
The F\"oppl von K\'arm\'an number measures the relative energy scales of stretching and of bending a sheet of size L; for reference, these simulation  values are similar to that of a standard 8.5'' $\times$ 11'' sheet of paper.  Thermal fluctuations become significant  beyond  the length-scale  \cite{Nelson_JPhys_1987,Kantor_PRA_1987}
 
\begin{equation}
\ell_{th} \approx \sqrt{\frac{16\pi^3 \kappa^2}{3Yk_BT}} \approx 0.4-5a.
\end{equation}
This is the scale at which the nonlinear stretching energy matches the bending energy; that is, the length-scale at which thermal corrections to the bending rigidity are comparable to the bare bending rigidity. 

To introduce a slit, we simply delete the edges and faces of the triangulation which cross the slit (see Fig.~\ref{Fig_triangulation}b).  This changes both energy terms; pairs of vertices joined by the deleted 
edges are removed from the stretching-energy sum, while dihedral angles involving deleted faces are removed from the bending-energy sum.  We consider slits which run along the length of the ribbon, centered both in width and in length, such that both the clamped and free ends are uncut.  We also consider a slit which extends to cut the free end, effectively giving two ribbons, 
each of half the total width, which are tethered together at the clamped end.

We simulate an initially flat ribbon, with a small random vertical displacement added to each unclamped vertex; this initial noise avoids the simulation getting trapped in the stationary state of the 
perfectly flat ribbon.  Simulations are carried out in an NVT ensemble using the HOOMD-Blue molecular dynamics toolkit \cite{HOOMD-Blue, Anderson_JCP_2008}, 
on a single GPU per simulation.  The unit of time in these simulations is $\tau \equiv \sqrt{\frac{ma^2}{k_BT}}$; the simulation time step used is $\delta t = 0.0025\tau$, and simulations are run 
for $2.5$-$5\times 10^6 \ \tau$; we carry out between 2 and 40 independent simulations for each ribbon configuration to improve our statistics. 
\begin{figure}
\centering
\includegraphics[width=1.0\columnwidth]{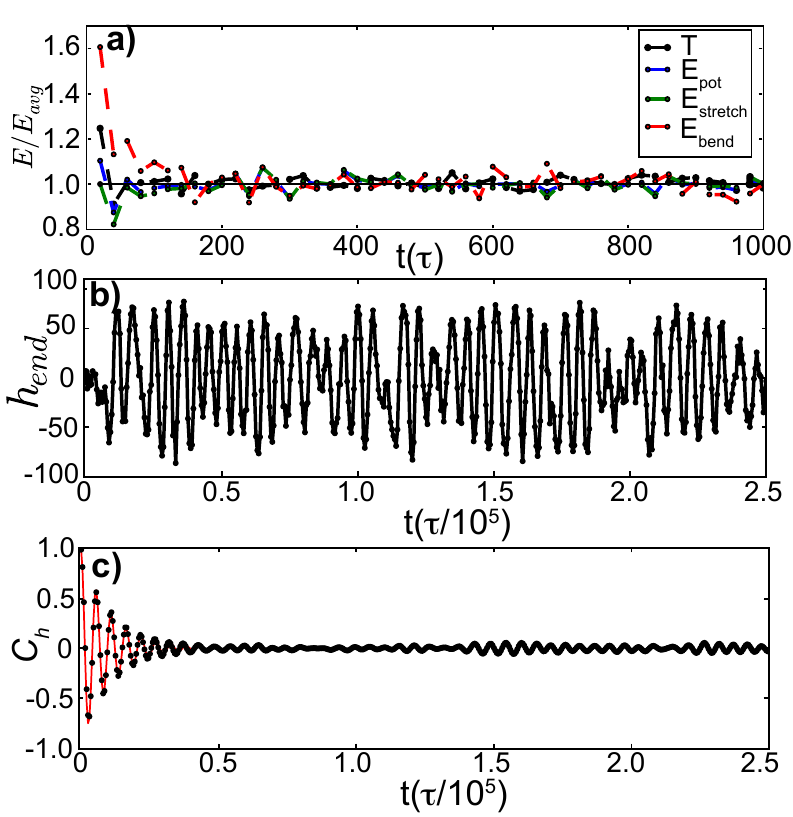}
\caption{Equilibration and relaxation times.  a) Evolution of the temperature and potential energy contributions in the first several $\times 10^5$ simulation timesteps, showing the decay of transient 
states present in the initial random configuration.  b) Evolution of the height of the free end of the ribbon in the first $1/20$ of a simulation, showing evidence of a thermally excited, long-lived oscillatory
 mode.  c) Time correlation function of the height of the free end, $C_h(t) = \frac{1}{t_{max}-t}\sum_{t^\prime}\left(h_{end}(t^\prime+t)-\bar{h}_{end}\right)\left(h_{end}(t^\prime)-\bar{h}_{end}\right)$ ($\bar{h}_{end}=\frac{1}{t_{max}}\sum_{t}h_{end}(t)\approx 0$ and we normalized the $C_h(t)$ to $1$ at $t=0$), 
 showing this long-lived mode and its decay; the red curve gives a fit of the data to the form $\exp{(-t/T_2)}\cos{(2\pi t/T_1)}$ with period $T_1=5.5\times10^3\tau$, and decay constant 
 $T_2 = 1.1\times10^4\tau$.  All panels show results for a ribbon with no slit; results for ribbons with slits are similar, with some increase in the relaxation times.}
\label{Fig_equilibration}
\end{figure}

\subsection{Equilibration and Relaxation}

A crucial consideration in molecular dynamics simulations of thermal systems is to determine the equilibration and relaxation times.  We distinguish two very different timescales in this system.

The `equilibration time' is the time it takes for the initial state with its added short-wavelength noise to decay to a typical state, so that the temperature and energy contributions reach their equilibrium values.  For this system, this is on the order of $250$-$500\tau$ (Fig.~\ref{Fig_equilibration}a).  We thus consider the first $500\tau$ as our equilibration period, and do not include the data from this period in our further analysis.

The `relaxation time' is the time in which thermally-excited modes decay, allowing for statistically independent sampling.  This relaxation time is much longer than the initial equilibration time, and may depend on the quantity being sampled.  We consider primarily the height of the free end of the ribbon, for which the relaxation time is $\approx 1.1\times10^4\tau$.  The ribbon is approximately described by the Euler-Bernoulli theory for a bending beam \cite{audoly2010elasticity}, for which the equations of motion admit periodic solutions; the lowest-energy such mode is thermally excited, as is apparent in the oscillations of the time trace of the height of the free end (Fig.~\ref{Fig_equilibration}b), and in a broad peak in the Fourier-transform of this time trace (data not shown).   
This mode decays quite slowly, as evident from the time correlation function shown in Fig.~\ref{Fig_equilibration}c.  Thus in order to obtain statistically independent samples of the end height, our sampling time must be similar to the relaxation time. We determined the optimal frequency of recording snapshots by calculating one estimate of the statistical error of the root mean-square height of the free end, $h^{end}_{rms}$, (defined below) by measuring $h^{end}_{rms}$ for each of many (up to 40) independent simulation runs, and taking the standard deviation of those independent measurements. For each sampling time, we then calculated an estimate of the error on $h^{end}_{rms}$ from the samples within one simulation and chose the sampling time for which these error estimates matched. Typically, the optimal sampling period was found to be $\approx 6\times10^3\tau$.

\section{Results}
\label{sec:results}

\subsection{Thermally Fluctuating Ribbons}

\begin{figure}
\centering
\includegraphics[width=\columnwidth]{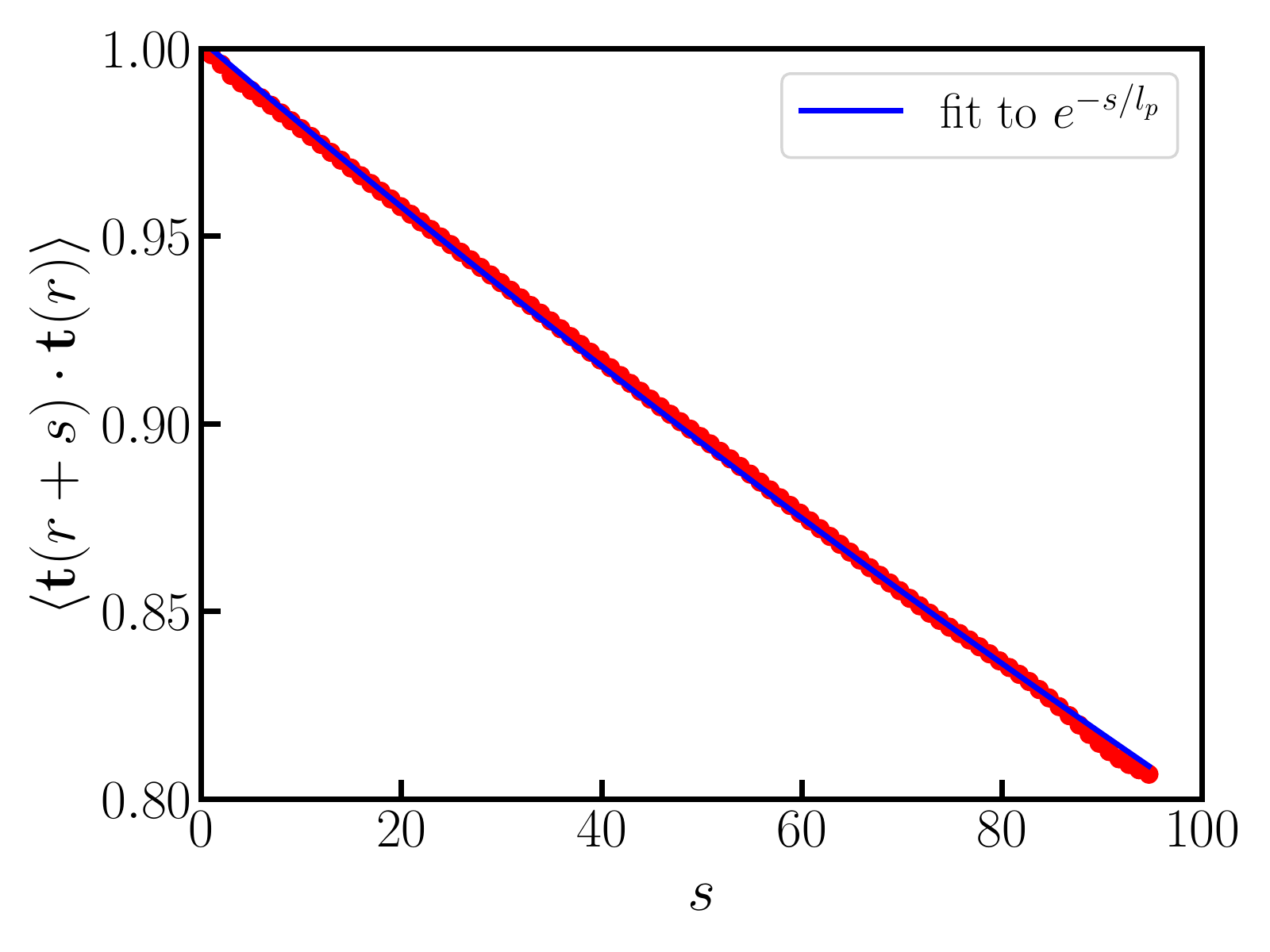}
\caption{Fit of the tangent-tangent correlation function, $\langle\mathbf{t}(s)\cdot\mathbf{t}(0)\rangle$ to an exponential function $e^{-s/\ell_p}$ for a ribbon of size $L=100a$, $W=10a$ and $\tilde{\kappa}=20k_BT$. Fitted $\ell_p\approx 440a$, while the simple analytic estimate for these parameters estimates it at $\approx350a$. }
\label{fig:perist_len}
\end{figure}

In most simulations we have chosen a bending rigidity $\tilde{\kappa}=5k_BT$. Thermal excitations cause the ribbons to fluctuate dramatically from their lowest energy, flat configuration -- several snapshots of these fluctuations are shown in Fig.~\ref{Fig_fluctuating_ribbons}, and videos are available in the Supplemental Material \cite{SM}. We estimate the persistence length $\ell_p$ of the ribbon by tracking the midline in the longitudinal direction of a ribbon with no slits and fitting its tangent-tangent correlation function to an exponential function: $\langle\mathbf{t}(s)\cdot\mathbf{t}(0)\rangle\propto e^{-s/\ell_p}$, where $s$ is the distance measured along the contour of the midline (Fig. \ref{fig:perist_len}). For this value of bending rigidity most measured values of $\ell_p$ were found to be comparable but slightly higher than the analytical estimate of $\ell_p=2W\kappa/k_BT=\sqrt{3}W\tilde{\kappa}/k_BT$. This means that in most cases the ribbon is in the semi-flexible regime $\ell_p \gtrsim L$, except for a very narrow ($W=5a$) ribbon, for which $\ell_p\approx L/2$.  For these parameters then, ribbons are soft enough to fluctuate significantly, sometimes even bending back on themselves, such as in Fig.~\ref{Fig_fluctuating_ribbons}c, but stiff enough  that these configurations are very rare; more frequently we see configurations such as in Fig.~\ref{Fig_fluctuating_ribbons}a,b.  It is important to point out that we do not observe self-intersection of the ribbons, although this is not explicitly prevented by the simulation model, which indicated that our simulations are far from the polymer-like limit. 

\begin{figure}
\centering
\includegraphics[width=\columnwidth]{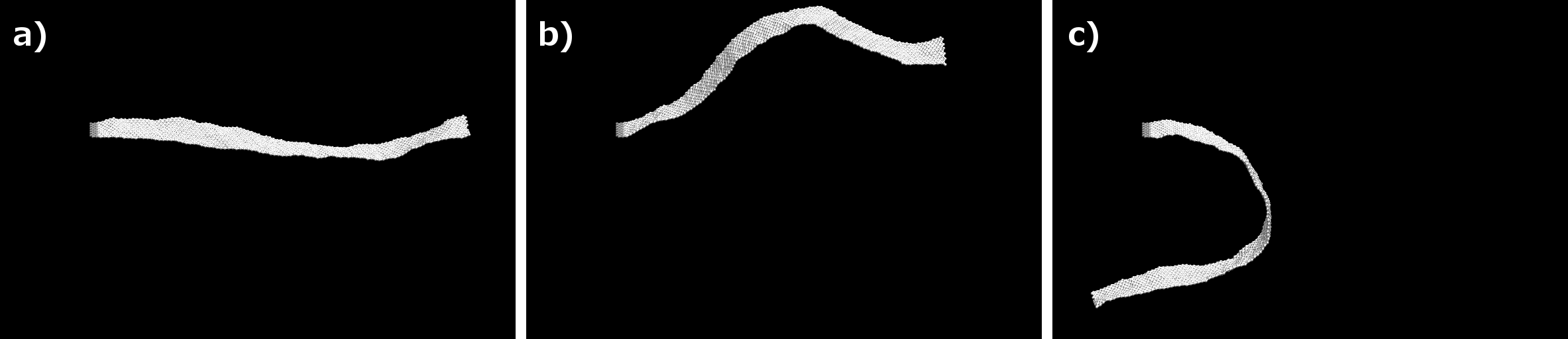}
\caption{Snapshots of a simulated ribbon undergoing thermal fluctuations.  The dark gray vertices are held fixed, `clamping' the leftmost end of the ribbon.  a) A configuration near flat; b) a moderate excursion; c) a large excursion in which the ribbon curves back on itself.}
\label{Fig_fluctuating_ribbons}
\end{figure}

We confirm that the average height of the free end, $h_{end}$, of the ribbon is consistent with zero, its expected equilibrium value.  We then quantify the scale of the fluctuations by calculating the root-mean-square (r.m.s.)~height $h_{rms}^{end}\equiv\sqrt{\langle h_{end}^2\rangle}$, where $\langle\dots\rangle$ denotes both the average over the width of the ribbon and over time, i.e., over statistically independent samples obtained from the molecular dynamics trajectory.  The r.m.s.~height is expected to scale as $h_{rms}^{end} \sim 1/\sqrt{\kappa_{eff}}$, where $\kappa_{eff}$ is an \emph{effective} bending rigidity.  These soft ribbons with no slits have an r.m.s.~height of $h_{rms}^{end} = 42.8a \pm 0.2a$.

\subsection{Suppression of Fluctuations by Slits}

\begin{figure*}
\centering
\includegraphics[width=0.95\textwidth]{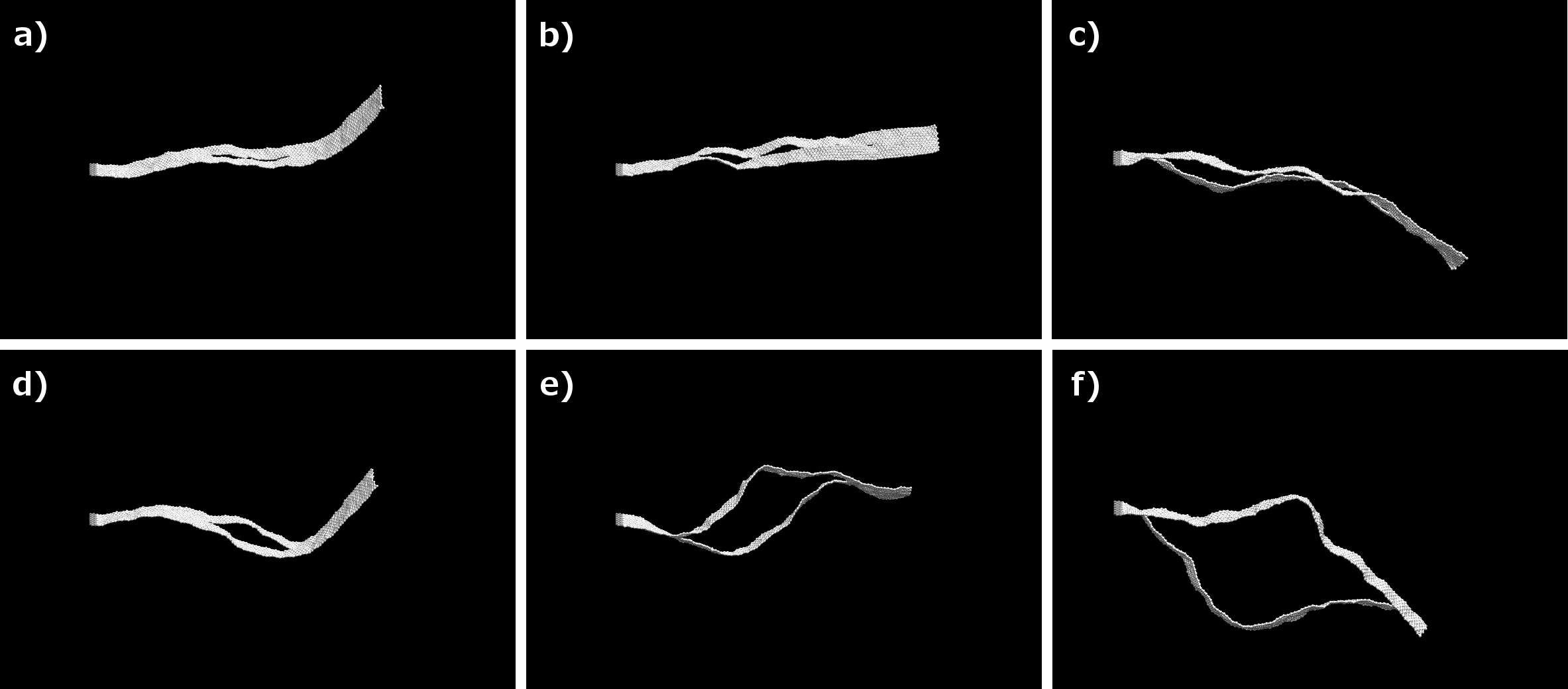}
\caption{Snapshots of simulated ribbons with slits undergoing thermal fluctuations.  The slits are centered in both dimensions, and extend 30\% (a,d); 60\% (b,e); and 90\% (c,f) of the total length of the ribbon.  In (a-c), the two sides of the ribbon remain close to one another, while in (d-f), a large gap is apparent as the two sides of the ribbon fluctuate in opposing directions.}
\label{Fig_fluctuating_slits}
\end{figure*}

A slit in the ribbon allows for new configurations, in which a gap opens between the two lips of the slit: snapshots of such configurations are shown in Fig.~\ref{Fig_fluctuating_slits} for several slit lengths.  Note that these gaps can become quite large.  We see occasional examples of the two sides of the ribbon intersecting (e.g.~in Fig.~\ref{Fig_fluctuating_slits}e), but such intersections are infrequent.

\begin{figure}
\centering
\includegraphics[width=0.95\columnwidth]{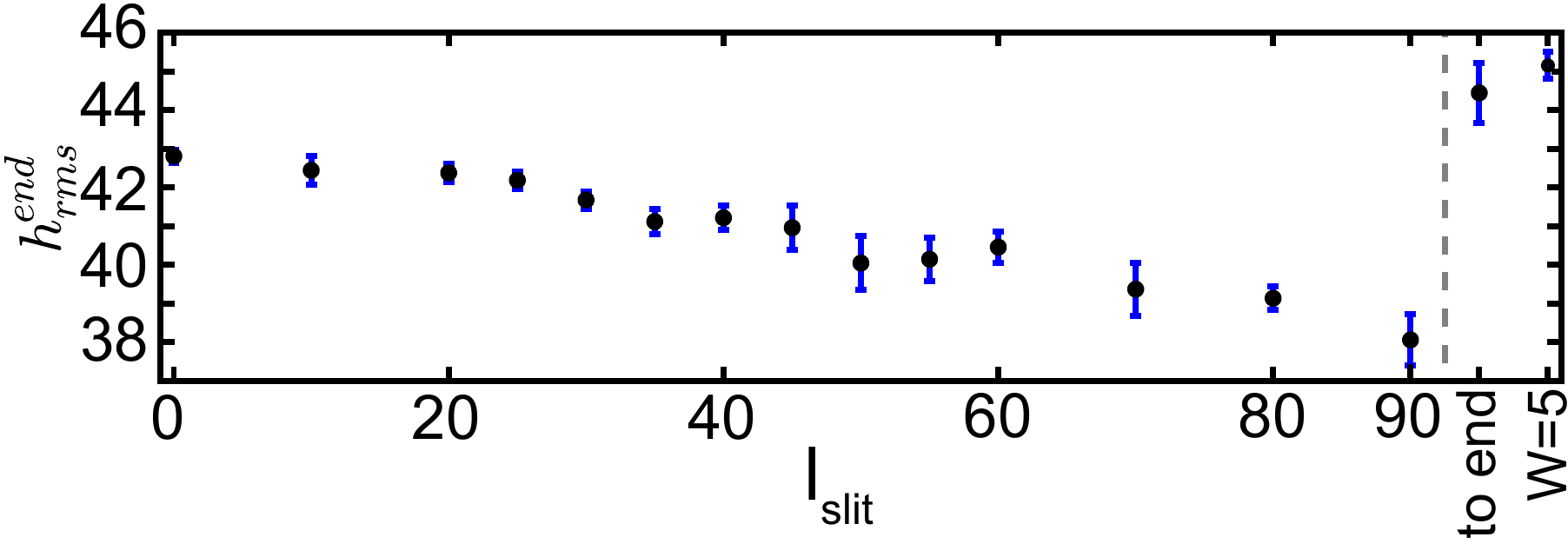}
\caption{Root-mean-square height, $h_{rms}^{end} $ of the free end of a simulated fluctuating ribbon as a function of slit length, $l_{slit}$.  Except for the rightmost two points, all ribbons have the same total length and width, and all slits are centered in both dimensions.  The point labeled `to end' represents a ribbon with a slit that is centered in width but extends through the free end of the ribbon; the point labeled `$W=5$' represents a single ribbon of the same total length but half the width, $W=5a$.  Note that the r.m.s.~height decreases as the slit length increases.}
\label{Fig_rms_heights}
\end{figure}

Strikingly, the presence of a slit significantly suppresses the overall amplitude of fluctuations of the ribbon.  We quantify the effect by the change in the r.m.s.~height of the free end of a ribbon; Fig.~\ref{Fig_rms_heights} shows the variation of $h_{rms}$ with the length of the slit.  It is clear that the presence of the slit \emph{reduces} the scale of the fluctuations, essentially \emph{stiffening} the ribbon to thermal fluctuations.  The effect is stronger for longer slits, with a maximum decrease of about 11\% to $h_{rms}^{end} = 38.1a \pm 0.7a$ for a slit extending 90\% of the length of the ribbon; this corresponds to an increase of about 20\% in the effective bending rigidity $\kappa_{eff}$.  Our results are suggestive that the effect is continuous, with the suppression present even for the shortest slits, with $l_{slit}=10a = W$; certainly the suppression is significant at the $4.0\sigma$ level for $l=30a$.  The effect disappears when the slit is extended through the free end, effectively resulting in two independent ribbons each of half the total width; indeed, in this case the behavior is consistent with that of a single ribbon of width $W=5a$, as shown by the rightmost two points in Fig.~\ref{Fig_fluctuating_slits}.  These thinner ribbons have slightly larger fluctuations than the ribbon of with $W=10a$, corresponding to a softer effective bending rigidity as predicted by the renormalization theory of \cite{Nelson_JPhys_1987}.

\begin{figure}
\centering
\includegraphics[width=0.95\columnwidth]{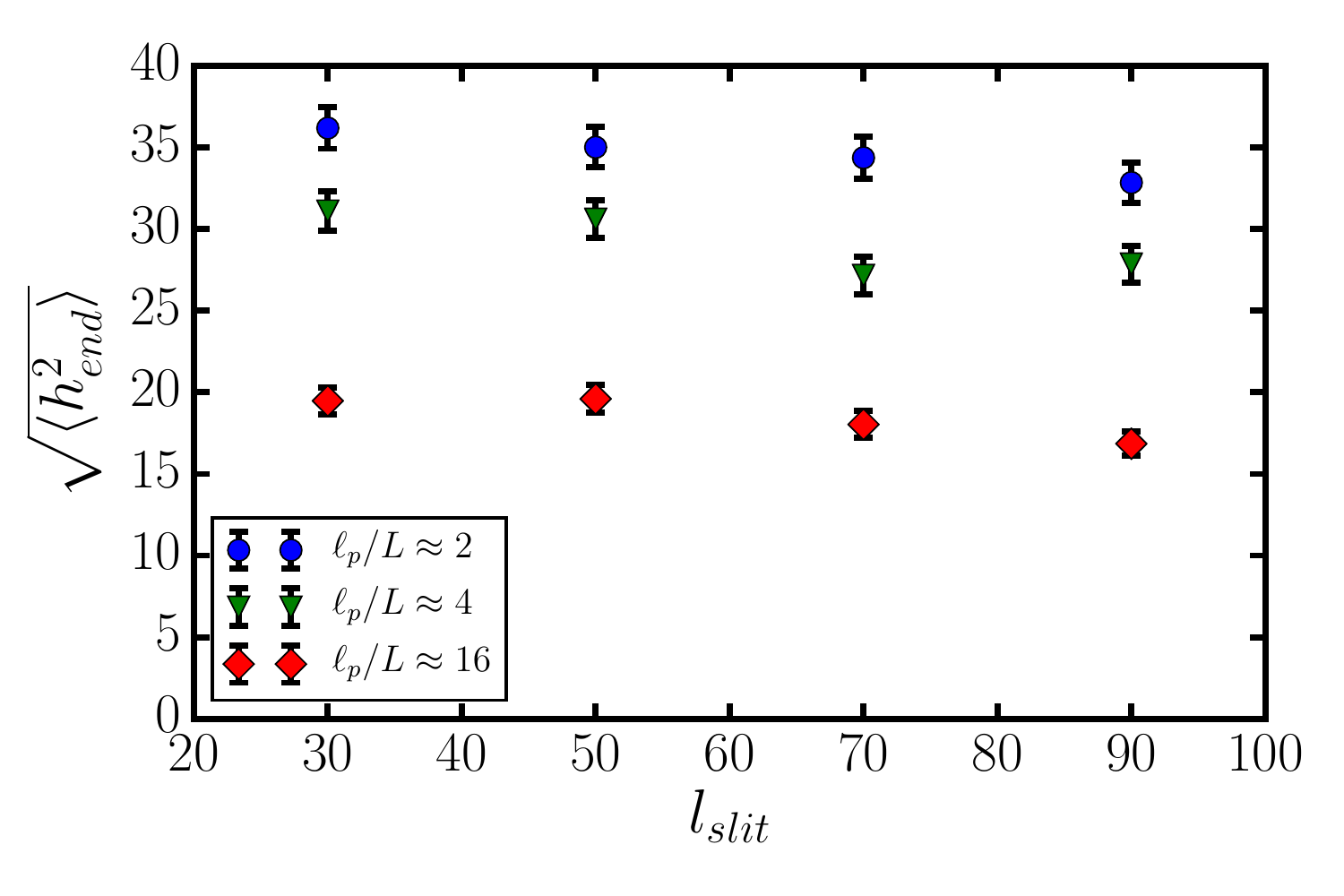}
\caption{Root-mean-square height, $h_{rms}^{end}\equiv\sqrt{\langle h_{end}^{2}\rangle}$ of free end of a ribbon of length $L=100a$ and width $W=10a$ and slit lengths $l_{slit}=30a, 50a, 70a, 90a$
for three different bending rigidities expressed in terms of the persistence length. Values of the persistence length $\ell_p$ were estimated using the $T=0$ analytical result: $\ell_p=2W\kappa/k_BT$.}
\label{fig:slits_kappa}
\end{figure}

In order to establish whether the effective stiffening of the ribbon by slits is affected by the bare bending stiffness, we extended our simulations deeper into the cantilever regime, $\ell_p >L$. This was achieved by simulating three larger values of the bare bending rigidity, $\tilde{\kappa} = 10$, 20 and $80 k_BT$, while keeping all other parameters fixed. The corresponding values of $\ell_p >L$ are 2, 4 and 16 respectively. We restricted this extended study to the $L=100a$, $W=10a$ case, with four different values of the slit length $l_{slit}=30a, 50a, 70a, 90a$. 
In Fig.~\ref{fig:slits_kappa} we show $h_{rms}^{end}$ as a function of the slit length. The overall trend in suppression of the r.m.s. height fluctuations of the free
end continues to hold in this entire regime which reaches a persistence length an order of magnitude larger than the system size.  We note that the regime $\ell_p\ll L$ is not the relevant one for our purposes, since the ribbon will effectively behave like a thickened polymer in this regime. Furthermore, simulations in the regime are ridden with many technical problems related to the ribbon being very soft and self-intersections being very common. 

\begin{figure}
\centering
\includegraphics[width=0.98\columnwidth]{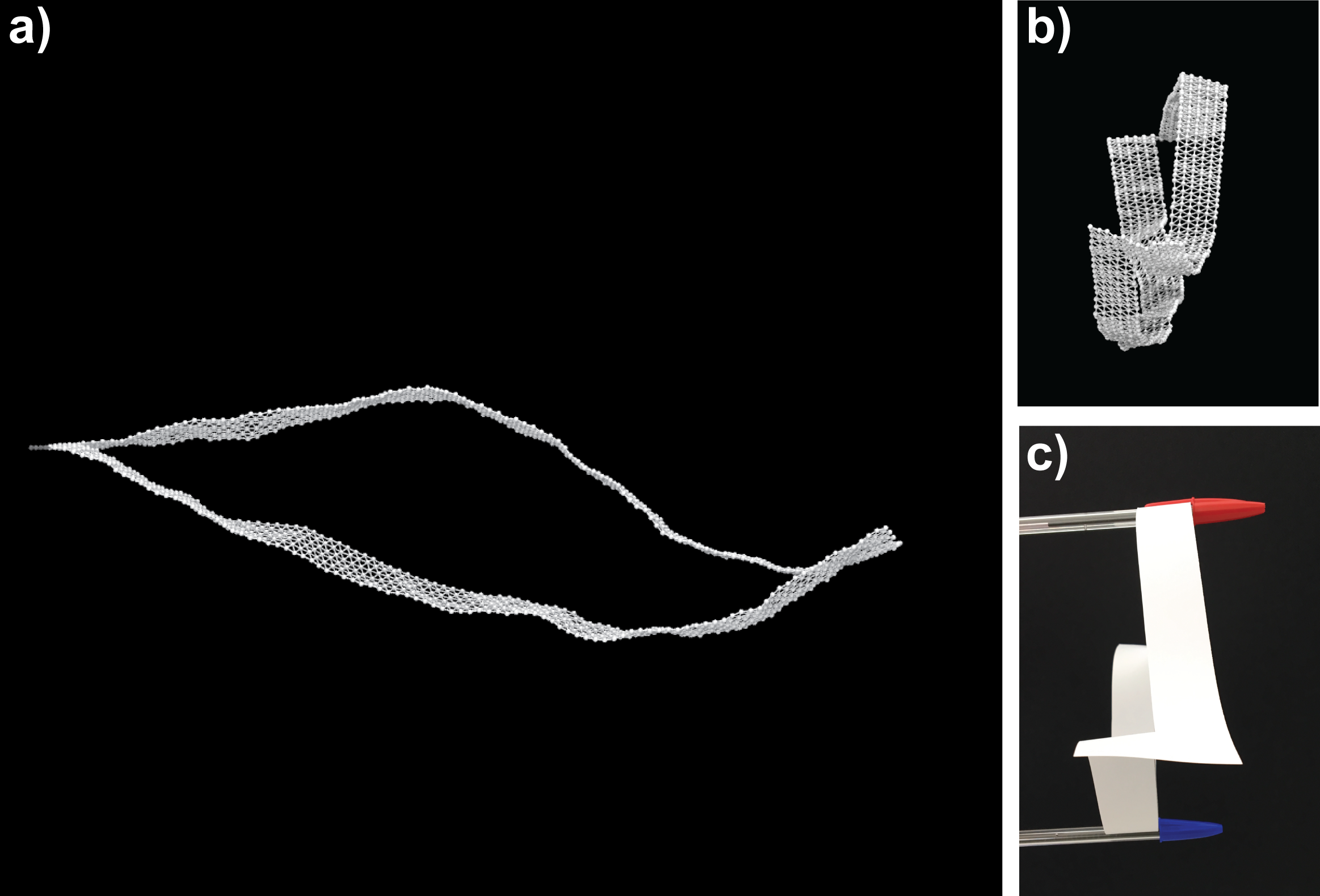}
\caption{A lip mode where two sides of the ribbon bend in opposite directions. Such configurations are 
often seen in the simulation: a) side view and b) front view. c) A similar configuration can be created with a
a narrow sheet of paper cut in the middle and two leaflets spread in opposite directions.}
\label{fig:lip_mode}
\end{figure}

One possible mechanism for suppressing height fluctuations is that a lip mode that is above the $z=0$ plane of the ribbon causes the half of the ribbon on the same side to bend below the $z=0$ (Fig.~\ref{fig:lip_mode}). Thus the height fluctuations will be lowered on average by this tendency to oppose. When one lip bends up and the other down (roughly $50\%$ of the time) this induces a down-up twist beyond the slit with a node at the midpoint. Twist in semiflexible ribbons has been studied in \cite{mergell2002statistical}. This canceling via bend-twist coupling increases as the slit length grows. We do not currently have an analytic derivation of this result {--} it is entirely conjectural based on considering the various modes of a thermally fluctuating ribbon \cite{footnote}. 

\section{Discussion}
\label{sec:discussion}

Our results demonstrate the importance of subtle changes in the geometry of a thermally fluctuating thin ribbon. The configuration space and the relative weights of allowed configurations are both sensitive to the presence of slits. The mechanical consequences of adding slits and holes and arrays of these modular elements will need to be fully understood before we can efficiently design meta-materials by the assembly of modules. Here we have shown that a simple slit in a thermally fluctuating ribbon leads to a marked stiffening of the effective bending rigidity in the longitudinal direction.  This non-intuitive effect is thermally driven and is yet another example of order from disorder. The effect for a single slit is moderate, with a change in the effective rigidity on the order of 10-20\%, but suitably designed more complex topologies are likely to yield larger effects, as in the kirigami springs reported in \cite{Blees_Nature_2015}.  

Many open questions remain, both about this simple system, and about the larger picture of controlling mechanical properties of 2D membranes using topology.  Further work on this system 
might determine the functional form of the bending rigidity as a function of slit length, and in particular, whether the stiffening effect is truly continuous, that is, present for an arbitrarily short slit, 
or whether it exhibits a threshold behavior, such that there is a non-zero minimum length below which a slit has no effect.  We have only considered the material properties in the direction longitudinal to the axis of the slit; experiments demonstrate that deformations in the \emph{transverse} direction can convert stretching energy to bending energy, dramatically softening the effective elastic modulus \cite{Blees_Nature_2015}.  This directional dependence on the effects of topology is still to be completely understood.  The experiments of \cite{Blees_Nature_2015} begin to explore the possibilities available using more complex arrangements of multiple slits and holes of different shapes; understanding more fully the interactions among different geometric elements will be essential to controlling the properties of 2D membranes.

The study of two-dimensional crystalline materials such as graphene as building blocks for nanoscale metamaterials is just beginning.  Simulations such as ours provide an essential complement 
to theory and experiment, providing insight into the ways in which we can control the properties of such materials using topology and geometry.

\section{Acknowledgements}
The authors would like to acknowledge the support and hospitality of the Institute for Computational and Experimental Research in Mathematics (ICERM) at Brown University, which hosted us and where ERR was a postdoctoral fellow while we were carrying out this research.  We thank David R.~Nelson, Andrej Ko{\v s}mrlj, Francesco Serafin and Suraj Shankar for invaluable discussions.  This research was conducted using computational resources at the Center for Computation and Visualization, Brown University.  Some figures and the supplementary videos were created using the Visual Molecular Dynamics (VMD) software \cite{VMD, Humphrey_JMG_1996}. The research of MJB was supported by the NSF DMREF grant 7454419. RS thanks the UK EPSRC (Grant No. EP/M009599/1) for funding.

\bibliographystyle{apsrev4-1}

\begin{thebibliography}{34}%
\makeatletter
\providecommand \@ifxundefined [1]{%
 \@ifx{#1\undefined}
}%
\providecommand \@ifnum [1]{%
 \ifnum #1\expandafter \@firstoftwo
 \else \expandafter \@secondoftwo
 \fi
}%
\providecommand \@ifx [1]{%
 \ifx #1\expandafter \@firstoftwo
 \else \expandafter \@secondoftwo
 \fi
}%
\providecommand \natexlab [1]{#1}%
\providecommand \enquote  [1]{``#1''}%
\providecommand \bibnamefont  [1]{#1}%
\providecommand \bibfnamefont [1]{#1}%
\providecommand \citenamefont [1]{#1}%
\providecommand \href@noop [0]{\@secondoftwo}%
\providecommand \href [0]{\begingroup \@sanitize@url \@href}%
\providecommand \@href[1]{\@@startlink{#1}\@@href}%
\providecommand \@@href[1]{\endgroup#1\@@endlink}%
\providecommand \@sanitize@url [0]{\catcode `\\12\catcode `\$12\catcode
  `\&12\catcode `\#12\catcode `\^12\catcode `\_12\catcode `\%12\relax}%
\providecommand \@@startlink[1]{}%
\providecommand \@@endlink[0]{}%
\providecommand \url  [0]{\begingroup\@sanitize@url \@url }%
\providecommand \@url [1]{\endgroup\@href {#1}{\urlprefix }}%
\providecommand \urlprefix  [0]{URL }%
\providecommand \Eprint [0]{\href }%
\providecommand \doibase [0]{http://dx.doi.org/}%
\providecommand \selectlanguage [0]{\@gobble}%
\providecommand \bibinfo  [0]{\@secondoftwo}%
\providecommand \bibfield  [0]{\@secondoftwo}%
\providecommand \translation [1]{[#1]}%
\providecommand \BibitemOpen [0]{}%
\providecommand \bibitemStop [0]{}%
\providecommand \bibitemNoStop [0]{.\EOS\space}%
\providecommand \EOS [0]{\spacefactor3000\relax}%
\providecommand \BibitemShut  [1]{\csname bibitem#1\endcsname}%
\let\auto@bib@innerbib\@empty
\bibitem [{\citenamefont {Novoselov}\ \emph {et~al.}(2005)\citenamefont
  {Novoselov}, \citenamefont {Geim}, \citenamefont {Morozov}, \citenamefont
  {Jiang}, \citenamefont {Katsnelson}, \citenamefont {Grigorieva},
  \citenamefont {Dubonos},\ and\ \citenamefont {Firsov}}]{novoselov2005two}%
  \BibitemOpen
  \bibfield  {author} {\bibinfo {author} {\bibfnamefont {K.~S.}\ \bibnamefont
  {Novoselov}}, \bibinfo {author} {\bibfnamefont {A.~K.}\ \bibnamefont {Geim}},
  \bibinfo {author} {\bibfnamefont {S.}~\bibnamefont {Morozov}}, \bibinfo
  {author} {\bibfnamefont {D.}~\bibnamefont {Jiang}}, \bibinfo {author}
  {\bibfnamefont {M.}~\bibnamefont {Katsnelson}}, \bibinfo {author}
  {\bibfnamefont {I.}~\bibnamefont {Grigorieva}}, \bibinfo {author}
  {\bibfnamefont {S.}~\bibnamefont {Dubonos}}, \ and\ \bibinfo {author}
  {\bibfnamefont {A.}~\bibnamefont {Firsov}},\ }\href@noop {} {\bibfield
  {journal} {\bibinfo  {journal} {nature}\ }\textbf {\bibinfo {volume} {438}},\
  \bibinfo {pages} {197} (\bibinfo {year} {2005})}\BibitemShut {NoStop}%
\bibitem [{\citenamefont {Wang}\ \emph {et~al.}(2012)\citenamefont {Wang},
  \citenamefont {Kalantar-Zadeh}, \citenamefont {Kis}, \citenamefont
  {Coleman},\ and\ \citenamefont {Strano}}]{wang2012electronics}%
  \BibitemOpen
  \bibfield  {author} {\bibinfo {author} {\bibfnamefont {Q.~H.}\ \bibnamefont
  {Wang}}, \bibinfo {author} {\bibfnamefont {K.}~\bibnamefont
  {Kalantar-Zadeh}}, \bibinfo {author} {\bibfnamefont {A.}~\bibnamefont {Kis}},
  \bibinfo {author} {\bibfnamefont {J.~N.}\ \bibnamefont {Coleman}}, \ and\
  \bibinfo {author} {\bibfnamefont {M.~S.}\ \bibnamefont {Strano}},\
  }\href@noop {} {\bibfield  {journal} {\bibinfo  {journal} {Nature
  nanotechnology}\ }\textbf {\bibinfo {volume} {7}},\ \bibinfo {pages} {699}
  (\bibinfo {year} {2012})}\BibitemShut {NoStop}%
\bibitem [{\citenamefont {Song}\ \emph {et~al.}(2010)\citenamefont {Song},
  \citenamefont {Ci}, \citenamefont {Lu}, \citenamefont {Sorokin},
  \citenamefont {Jin}, \citenamefont {Ni}, \citenamefont {Kvashnin},
  \citenamefont {Kvashnin}, \citenamefont {Lou}, \citenamefont {Yakobson} \emph
  {et~al.}}]{song2010large}%
  \BibitemOpen
  \bibfield  {author} {\bibinfo {author} {\bibfnamefont {L.}~\bibnamefont
  {Song}}, \bibinfo {author} {\bibfnamefont {L.}~\bibnamefont {Ci}}, \bibinfo
  {author} {\bibfnamefont {H.}~\bibnamefont {Lu}}, \bibinfo {author}
  {\bibfnamefont {P.~B.}\ \bibnamefont {Sorokin}}, \bibinfo {author}
  {\bibfnamefont {C.}~\bibnamefont {Jin}}, \bibinfo {author} {\bibfnamefont
  {J.}~\bibnamefont {Ni}}, \bibinfo {author} {\bibfnamefont {A.~G.}\
  \bibnamefont {Kvashnin}}, \bibinfo {author} {\bibfnamefont {D.~G.}\
  \bibnamefont {Kvashnin}}, \bibinfo {author} {\bibfnamefont {J.}~\bibnamefont
  {Lou}}, \bibinfo {author} {\bibfnamefont {B.~I.}\ \bibnamefont {Yakobson}},
  \emph {et~al.},\ }\href@noop {} {\bibfield  {journal} {\bibinfo  {journal}
  {Nano letters}\ }\textbf {\bibinfo {volume} {10}},\ \bibinfo {pages} {3209}
  (\bibinfo {year} {2010})}\BibitemShut {NoStop}%
\bibitem [{\citenamefont {Li}\ \emph {et~al.}(2014)\citenamefont {Li},
  \citenamefont {Yu}, \citenamefont {Ye}, \citenamefont {Ge}, \citenamefont
  {Ou}, \citenamefont {Wu}, \citenamefont {Feng}, \citenamefont {Chen},\ and\
  \citenamefont {Zhang}}]{li2014black}%
  \BibitemOpen
  \bibfield  {author} {\bibinfo {author} {\bibfnamefont {L.}~\bibnamefont
  {Li}}, \bibinfo {author} {\bibfnamefont {Y.}~\bibnamefont {Yu}}, \bibinfo
  {author} {\bibfnamefont {G.~J.}\ \bibnamefont {Ye}}, \bibinfo {author}
  {\bibfnamefont {Q.}~\bibnamefont {Ge}}, \bibinfo {author} {\bibfnamefont
  {X.}~\bibnamefont {Ou}}, \bibinfo {author} {\bibfnamefont {H.}~\bibnamefont
  {Wu}}, \bibinfo {author} {\bibfnamefont {D.}~\bibnamefont {Feng}}, \bibinfo
  {author} {\bibfnamefont {X.~H.}\ \bibnamefont {Chen}}, \ and\ \bibinfo
  {author} {\bibfnamefont {Y.}~\bibnamefont {Zhang}},\ }\href@noop {}
  {\bibfield  {journal} {\bibinfo  {journal} {Nature nanotechnology}\ }\textbf
  {\bibinfo {volume} {9}},\ \bibinfo {pages} {372} (\bibinfo {year}
  {2014})}\BibitemShut {NoStop}%
\bibitem [{\citenamefont {Geim}\ and\ \citenamefont
  {Grigorieva}(2013)}]{geim2013van}%
  \BibitemOpen
  \bibfield  {author} {\bibinfo {author} {\bibfnamefont {A.~K.}\ \bibnamefont
  {Geim}}\ and\ \bibinfo {author} {\bibfnamefont {I.~V.}\ \bibnamefont
  {Grigorieva}},\ }\href@noop {} {\bibfield  {journal} {\bibinfo  {journal}
  {Nature}\ }\textbf {\bibinfo {volume} {499}},\ \bibinfo {pages} {419}
  (\bibinfo {year} {2013})}\BibitemShut {NoStop}%
\bibitem [{\citenamefont {Min}\ and\ \citenamefont
  {Aluru}(2011)}]{min2011mechanical}%
  \BibitemOpen
  \bibfield  {author} {\bibinfo {author} {\bibfnamefont {K.}~\bibnamefont
  {Min}}\ and\ \bibinfo {author} {\bibfnamefont {N.}~\bibnamefont {Aluru}},\
  }\href@noop {} {\bibfield  {journal} {\bibinfo  {journal}
  {Appl.~Phys.~Lett.}\ }\textbf {\bibinfo {volume} {98}},\ \bibinfo {pages}
  {013113} (\bibinfo {year} {2011})}\BibitemShut {NoStop}%
\bibitem [{\citenamefont {Ovid'ko}(2013)}]{ovid2013mechanical}%
  \BibitemOpen
  \bibfield  {author} {\bibinfo {author} {\bibfnamefont {I.}~\bibnamefont
  {Ovid'ko}},\ }\href@noop {} {\bibfield  {journal} {\bibinfo  {journal}
  {Rev.~Adv.~Mater.~Sci}\ }\textbf {\bibinfo {volume} {34}},\ \bibinfo {pages}
  {1} (\bibinfo {year} {2013})}\BibitemShut {NoStop}%
\bibitem [{\citenamefont {Blees}\ \emph {et~al.}(2015)\citenamefont {Blees},
  \citenamefont {Barnard}, \citenamefont {Rose}, \citenamefont {Roberts},
  \citenamefont {McGill}, \citenamefont {Huang}, \citenamefont {Ruyack},
  \citenamefont {Kevek}, \citenamefont {Kobrin}, \citenamefont {Muller},\ and\
  \citenamefont {McEuen}}]{Blees_Nature_2015}%
  \BibitemOpen
  \bibfield  {author} {\bibinfo {author} {\bibfnamefont {M.~K.}\ \bibnamefont
  {Blees}}, \bibinfo {author} {\bibfnamefont {A.~W.}\ \bibnamefont {Barnard}},
  \bibinfo {author} {\bibfnamefont {P.~A.}\ \bibnamefont {Rose}}, \bibinfo
  {author} {\bibfnamefont {S.~P.}\ \bibnamefont {Roberts}}, \bibinfo {author}
  {\bibfnamefont {K.~L.}\ \bibnamefont {McGill}}, \bibinfo {author}
  {\bibfnamefont {P.~Y.}\ \bibnamefont {Huang}}, \bibinfo {author}
  {\bibfnamefont {A.~R.}\ \bibnamefont {Ruyack}}, \bibinfo {author}
  {\bibfnamefont {J.~W.}\ \bibnamefont {Kevek}}, \bibinfo {author}
  {\bibfnamefont {B.}~\bibnamefont {Kobrin}}, \bibinfo {author} {\bibfnamefont
  {D.~A.}\ \bibnamefont {Muller}}, \ and\ \bibinfo {author} {\bibfnamefont
  {P.~L.}\ \bibnamefont {McEuen}},\ }\href {\doibase 10.1038/nature14588}
  {\bibfield  {journal} {\bibinfo  {journal} {Nature}\ } (\bibinfo {year}
  {2015}),\ 10.1038/nature14588},\ \bibinfo {note} {dOI
  10.1038/nature14588}\BibitemShut {NoStop}%
\bibitem [{\citenamefont {Nelson}\ and\ \citenamefont
  {Peliti}(1987)}]{Nelson_JPhys_1987}%
  \BibitemOpen
  \bibfield  {author} {\bibinfo {author} {\bibfnamefont {D.}~\bibnamefont
  {Nelson}}\ and\ \bibinfo {author} {\bibfnamefont {L.}~\bibnamefont
  {Peliti}},\ }\href {\doibase 10.1051/jphys:019870048070108500} {\bibfield
  {journal} {\bibinfo  {journal} {J.~Physique}\ }\textbf {\bibinfo {volume}
  {48}},\ \bibinfo {pages} {1085} (\bibinfo {year} {1987})}\BibitemShut
  {NoStop}%
\bibitem [{\citenamefont {Aronovitz}\ and\ \citenamefont
  {Lubensky}(1988)}]{aronovitz1988fluctuations}%
  \BibitemOpen
  \bibfield  {author} {\bibinfo {author} {\bibfnamefont {J.~A.}\ \bibnamefont
  {Aronovitz}}\ and\ \bibinfo {author} {\bibfnamefont {T.~C.}\ \bibnamefont
  {Lubensky}},\ }\href@noop {} {\bibfield  {journal} {\bibinfo  {journal}
  {Phys.~Rev.~Lett.}\ }\textbf {\bibinfo {volume} {60}},\ \bibinfo {pages}
  {2634} (\bibinfo {year} {1988})}\BibitemShut {NoStop}%
\bibitem [{\citenamefont {Nelson}\ \emph {et~al.}(2004)\citenamefont {Nelson},
  \citenamefont {Piran},\ and\ \citenamefont
  {Weinberg}}]{nelson2004statistical}%
  \BibitemOpen
  \bibfield  {author} {\bibinfo {author} {\bibfnamefont {D.}~\bibnamefont
  {Nelson}}, \bibinfo {author} {\bibfnamefont {T.}~\bibnamefont {Piran}}, \
  and\ \bibinfo {author} {\bibfnamefont {S.}~\bibnamefont {Weinberg}},\
  }\href@noop {} {\emph {\bibinfo {title} {Statistical mechanics of membranes
  and surfaces}}}\ (\bibinfo  {publisher} {World Scientific},\ \bibinfo {year}
  {2004})\BibitemShut {NoStop}%
\bibitem [{\citenamefont {Bowick}\ and\ \citenamefont
  {Travesset}(2001)}]{Bowick_PhysRep_2001}%
  \BibitemOpen
  \bibfield  {author} {\bibinfo {author} {\bibfnamefont {M.~J.}\ \bibnamefont
  {Bowick}}\ and\ \bibinfo {author} {\bibfnamefont {A.}~\bibnamefont
  {Travesset}},\ }\href {\doibase 10.1016/S0370-1573(00)00128-9} {\bibfield
  {journal} {\bibinfo  {journal} {Phys.~Rep.}\ }\textbf {\bibinfo {volume}
  {344}},\ \bibinfo {pages} {255} (\bibinfo {year} {2001})}\BibitemShut
  {NoStop}%
\bibitem [{\citenamefont {Paulose}\ \emph {et~al.}(2012)\citenamefont
  {Paulose}, \citenamefont {Vliegenthart}, \citenamefont {Gompper},\ and\
  \citenamefont {Nelson}}]{paulose2012fluctuating}%
  \BibitemOpen
  \bibfield  {author} {\bibinfo {author} {\bibfnamefont {J.}~\bibnamefont
  {Paulose}}, \bibinfo {author} {\bibfnamefont {G.~A.}\ \bibnamefont
  {Vliegenthart}}, \bibinfo {author} {\bibfnamefont {G.}~\bibnamefont
  {Gompper}}, \ and\ \bibinfo {author} {\bibfnamefont {D.~R.}\ \bibnamefont
  {Nelson}},\ }\href@noop {} {\bibfield  {journal} {\bibinfo  {journal}
  {Proc.~Natl.~Acad.~Sci.}\ }\textbf {\bibinfo {volume} {109}},\ \bibinfo
  {pages} {19551} (\bibinfo {year} {2012})}\BibitemShut {NoStop}%
\bibitem [{\citenamefont {Kantor}\ and\ \citenamefont
  {Nelson}(1987)}]{Kantor_PRA_1987}%
  \BibitemOpen
  \bibfield  {author} {\bibinfo {author} {\bibfnamefont {Y.}~\bibnamefont
  {Kantor}}\ and\ \bibinfo {author} {\bibfnamefont {D.~R.}\ \bibnamefont
  {Nelson}},\ }\href {\doibase 10.1103/PhysRevA.36.4020} {\bibfield  {journal}
  {\bibinfo  {journal} {Phys. Rev. A}\ }\textbf {\bibinfo {volume} {36}},\
  \bibinfo {pages} {4020} (\bibinfo {year} {1987})}\BibitemShut {NoStop}%
\bibitem [{\citenamefont {Le~Doussal}\ and\ \citenamefont
  {Radzihovsky}(1992)}]{le1992self}%
  \BibitemOpen
  \bibfield  {author} {\bibinfo {author} {\bibfnamefont {P.}~\bibnamefont
  {Le~Doussal}}\ and\ \bibinfo {author} {\bibfnamefont {L.}~\bibnamefont
  {Radzihovsky}},\ }\href@noop {} {\bibfield  {journal} {\bibinfo  {journal}
  {Phys.~Rev.~Lett.}\ }\textbf {\bibinfo {volume} {69}},\ \bibinfo {pages}
  {1209} (\bibinfo {year} {1992})}\BibitemShut {NoStop}%
\bibitem [{\citenamefont {Gazit}(2009)}]{gazit2009structure}%
  \BibitemOpen
  \bibfield  {author} {\bibinfo {author} {\bibfnamefont {D.}~\bibnamefont
  {Gazit}},\ }\href@noop {} {\bibfield  {journal} {\bibinfo  {journal}
  {Phys.~Rev.~E}\ }\textbf {\bibinfo {volume} {80}},\ \bibinfo {pages} {041117}
  (\bibinfo {year} {2009})}\BibitemShut {NoStop}%
\bibitem [{\citenamefont {Kownacki}\ and\ \citenamefont
  {Mouhanna}(2009)}]{kownacki2009crumpling}%
  \BibitemOpen
  \bibfield  {author} {\bibinfo {author} {\bibfnamefont {J.-P.}\ \bibnamefont
  {Kownacki}}\ and\ \bibinfo {author} {\bibfnamefont {D.}~\bibnamefont
  {Mouhanna}},\ }\href@noop {} {\bibfield  {journal} {\bibinfo  {journal}
  {Phys.~Rev.~E}\ }\textbf {\bibinfo {volume} {79}},\ \bibinfo {pages} {040101}
  (\bibinfo {year} {2009})}\BibitemShut {NoStop}%
\bibitem [{\citenamefont {Braghin}\ and\ \citenamefont
  {Hasselmann}(2010)}]{braghin2010thermal}%
  \BibitemOpen
  \bibfield  {author} {\bibinfo {author} {\bibfnamefont {F.}~\bibnamefont
  {Braghin}}\ and\ \bibinfo {author} {\bibfnamefont {N.}~\bibnamefont
  {Hasselmann}},\ }\href@noop {} {\bibfield  {journal} {\bibinfo  {journal}
  {Phys.~Rev.~B}\ }\textbf {\bibinfo {volume} {82}},\ \bibinfo {pages} {035407}
  (\bibinfo {year} {2010})}\BibitemShut {NoStop}%
\bibitem [{\citenamefont {Tr{\"o}ster}(2013)}]{troster2013high}%
  \BibitemOpen
  \bibfield  {author} {\bibinfo {author} {\bibfnamefont {A.}~\bibnamefont
  {Tr{\"o}ster}},\ }\href@noop {} {\bibfield  {journal} {\bibinfo  {journal}
  {Phy.~Rev.~B}\ }\textbf {\bibinfo {volume} {87}},\ \bibinfo {pages} {104112}
  (\bibinfo {year} {2013})}\BibitemShut {NoStop}%
\bibitem [{\citenamefont {Tr{\"o}ster}(2015)}]{troster2015fourier}%
  \BibitemOpen
  \bibfield  {author} {\bibinfo {author} {\bibfnamefont {A.}~\bibnamefont
  {Tr{\"o}ster}},\ }\href@noop {} {\bibfield  {journal} {\bibinfo  {journal}
  {Phys.~Rev.~E}\ }\textbf {\bibinfo {volume} {91}},\ \bibinfo {pages} {022132}
  (\bibinfo {year} {2015})}\BibitemShut {NoStop}%
\bibitem [{\citenamefont {Bowick}\ \emph {et~al.}(1996)\citenamefont {Bowick},
  \citenamefont {Catterall}, \citenamefont {Falcioni}, \citenamefont
  {Thorleifsson},\ and\ \citenamefont {Anagnostopoulos}}]{bowick1996flat}%
  \BibitemOpen
  \bibfield  {author} {\bibinfo {author} {\bibfnamefont {M.~J.}\ \bibnamefont
  {Bowick}}, \bibinfo {author} {\bibfnamefont {S.~M.}\ \bibnamefont
  {Catterall}}, \bibinfo {author} {\bibfnamefont {M.}~\bibnamefont {Falcioni}},
  \bibinfo {author} {\bibfnamefont {G.}~\bibnamefont {Thorleifsson}}, \ and\
  \bibinfo {author} {\bibfnamefont {K.~N.}\ \bibnamefont {Anagnostopoulos}},\
  }\href@noop {} {\bibfield  {journal} {\bibinfo  {journal} {J.~Phys.~I}\
  }\textbf {\bibinfo {volume} {6}},\ \bibinfo {pages} {1321} (\bibinfo {year}
  {1996})}\BibitemShut {NoStop}%
\bibitem [{\citenamefont {Los}\ \emph {et~al.}(2009)\citenamefont {Los},
  \citenamefont {Katsnelson}, \citenamefont {Yazyev}, \citenamefont
  {Zakharchenko},\ and\ \citenamefont {Fasolino}}]{los2009scaling}%
  \BibitemOpen
  \bibfield  {author} {\bibinfo {author} {\bibfnamefont {J.}~\bibnamefont
  {Los}}, \bibinfo {author} {\bibfnamefont {M.~I.}\ \bibnamefont {Katsnelson}},
  \bibinfo {author} {\bibfnamefont {O.}~\bibnamefont {Yazyev}}, \bibinfo
  {author} {\bibfnamefont {K.}~\bibnamefont {Zakharchenko}}, \ and\ \bibinfo
  {author} {\bibfnamefont {A.}~\bibnamefont {Fasolino}},\ }\href@noop {}
  {\bibfield  {journal} {\bibinfo  {journal} {Phys.~Rev.~B}\ }\textbf {\bibinfo
  {volume} {80}},\ \bibinfo {pages} {121405} (\bibinfo {year}
  {2009})}\BibitemShut {NoStop}%
\bibitem [{\citenamefont {Yllanes}\ \emph {et~al.}(2017)\citenamefont
  {Yllanes}, \citenamefont {Bhabesh}, \citenamefont {Nelson},\ and\
  \citenamefont {Bowick}}]{yllanes2017thermal}%
  \BibitemOpen
  \bibfield  {author} {\bibinfo {author} {\bibfnamefont {D.}~\bibnamefont
  {Yllanes}}, \bibinfo {author} {\bibfnamefont {S.}~\bibnamefont {Bhabesh}},
  \bibinfo {author} {\bibfnamefont {D.~R.}\ \bibnamefont {Nelson}}, \ and\
  \bibinfo {author} {\bibfnamefont {M.~J.}\ \bibnamefont {Bowick}},\
  }\href@noop {} {\bibfield  {journal} {\bibinfo  {journal} {arXiv e-prints}\ }
  (\bibinfo {year} {2017})},\ \Eprint {http://arxiv.org/abs/1705.07379}
  {1705.07379} \BibitemShut {NoStop}%
\bibitem [{\citenamefont {Bowick}\ \emph {et~al.}(2017)\citenamefont {Bowick},
  \citenamefont {Ko{\v{s}}mrlj}, \citenamefont {Nelson},\ and\ \citenamefont
  {Sknepnek}}]{bowick2017non}%
  \BibitemOpen
  \bibfield  {author} {\bibinfo {author} {\bibfnamefont {M.~J.}\ \bibnamefont
  {Bowick}}, \bibinfo {author} {\bibfnamefont {A.}~\bibnamefont
  {Ko{\v{s}}mrlj}}, \bibinfo {author} {\bibfnamefont {D.~R.}\ \bibnamefont
  {Nelson}}, \ and\ \bibinfo {author} {\bibfnamefont {R.}~\bibnamefont
  {Sknepnek}},\ }\href@noop {} {\bibfield  {journal} {\bibinfo  {journal}
  {Phys.~Rev.~B}\ }\textbf {\bibinfo {volume} {95}},\ \bibinfo {pages} {104109}
  (\bibinfo {year} {2017})}\BibitemShut {NoStop}%
\bibitem [{\citenamefont {Seung}\ and\ \citenamefont
  {Nelson}(1988)}]{Seung_PRA_1988}%
  \BibitemOpen
  \bibfield  {author} {\bibinfo {author} {\bibfnamefont {H.~S.}\ \bibnamefont
  {Seung}}\ and\ \bibinfo {author} {\bibfnamefont {D.~R.}\ \bibnamefont
  {Nelson}},\ }\href {\doibase 10.1103/PhysRevA.38.1005} {\bibfield  {journal}
  {\bibinfo  {journal} {Phys.~Rev.~A}\ }\textbf {\bibinfo {volume} {38}},\
  \bibinfo {pages} {1005} (\bibinfo {year} {1988})}\BibitemShut {NoStop}%
\bibitem [{\citenamefont {Schmidt}\ and\ \citenamefont
  {Fraternali}(2012)}]{Schmidt_JMPS_2012}%
  \BibitemOpen
  \bibfield  {author} {\bibinfo {author} {\bibfnamefont {B.}~\bibnamefont
  {Schmidt}}\ and\ \bibinfo {author} {\bibfnamefont {F.}~\bibnamefont
  {Fraternali}},\ }\href {\doibase 10.1016/j.jmps.2011.09.003} {\bibfield
  {journal} {\bibinfo  {journal} {J.~Mech.~Phys.~Solids}\ }\textbf {\bibinfo
  {volume} {60}},\ \bibinfo {pages} {172} (\bibinfo {year} {2012})}\BibitemShut
  {NoStop}%
\bibitem [{HOO()}]{HOOMD-Blue}%
  \BibitemOpen
  \href@noop {} {}\bibinfo {note}
  {http://codeblue.umich.edu/hoomd-blue}\BibitemShut {NoStop}%
\bibitem [{\citenamefont {Anderson}\ \emph {et~al.}(2008)\citenamefont
  {Anderson}, \citenamefont {Lorenz},\ and\ \citenamefont
  {Travesset}}]{Anderson_JCP_2008}%
  \BibitemOpen
  \bibfield  {author} {\bibinfo {author} {\bibfnamefont {J.~A.}\ \bibnamefont
  {Anderson}}, \bibinfo {author} {\bibfnamefont {C.~D.}\ \bibnamefont
  {Lorenz}}, \ and\ \bibinfo {author} {\bibfnamefont {A.}~\bibnamefont
  {Travesset}},\ }\href {\doibase 10.1016/j.jcp.2008.01.047} {\bibfield
  {journal} {\bibinfo  {journal} {J.~Comp.~Phys.}\ }\textbf {\bibinfo {volume}
  {227}},\ \bibinfo {pages} {5342} (\bibinfo {year} {2008})}\BibitemShut
  {NoStop}%
\bibitem [{\citenamefont {Audoly}\ and\ \citenamefont
  {Pomeau}(2010)}]{audoly2010elasticity}%
  \BibitemOpen
  \bibfield  {author} {\bibinfo {author} {\bibfnamefont {B.}~\bibnamefont
  {Audoly}}\ and\ \bibinfo {author} {\bibfnamefont {Y.}~\bibnamefont
  {Pomeau}},\ }\href@noop {} {\emph {\bibinfo {title} {Elasticity and geometry:
  from hair curls to the non-linear response of shells}}}\ (\bibinfo
  {publisher} {Oxford University Press},\ \bibinfo {year} {2010})\BibitemShut
  {NoStop}%
\bibitem [{SM()}]{SM}%
  \BibitemOpen
  \href@noop {} {}\bibinfo {note} {See Supplemental Material at [URL will be
  inserted by publisher].}\BibitemShut {Stop}%
\bibitem [{\citenamefont {Mergell}\ \emph {et~al.}(2002)\citenamefont
  {Mergell}, \citenamefont {Ejtehadi},\ and\ \citenamefont
  {Everaers}}]{mergell2002statistical}%
  \BibitemOpen
  \bibfield  {author} {\bibinfo {author} {\bibfnamefont {B.}~\bibnamefont
  {Mergell}}, \bibinfo {author} {\bibfnamefont {M.~R.}\ \bibnamefont
  {Ejtehadi}}, \ and\ \bibinfo {author} {\bibfnamefont {R.}~\bibnamefont
  {Everaers}},\ }\href@noop {} {\bibfield  {journal} {\bibinfo  {journal}
  {Phys.~Rev.~E}\ }\textbf {\bibinfo {volume} {66}},\ \bibinfo {pages} {011903}
  (\bibinfo {year} {2002})}\BibitemShut {NoStop}%
\bibitem [{foo()}]{footnote}%
  \BibitemOpen
  \href@noop {} {}\bibinfo {note} {We thank the referee for suggesting that we
  think about twist at the end of the ribbon.}\BibitemShut {Stop}%
\bibitem [{VMD()}]{VMD}%
  \BibitemOpen
  \href@noop {} {}\bibinfo {note}
  {http://www.ks.uiuc.edu/Research/vmd/}\BibitemShut {NoStop}%
\bibitem [{\citenamefont {Humphrey}\ \emph {et~al.}(1996)\citenamefont
  {Humphrey}, \citenamefont {Dalke},\ and\ \citenamefont
  {Schulten}}]{Humphrey_JMG_1996}%
  \BibitemOpen
  \bibfield  {author} {\bibinfo {author} {\bibfnamefont {W.}~\bibnamefont
  {Humphrey}}, \bibinfo {author} {\bibfnamefont {A.}~\bibnamefont {Dalke}}, \
  and\ \bibinfo {author} {\bibfnamefont {K.}~\bibnamefont {Schulten}},\
  }\href@noop {} {\bibfield  {journal} {\bibinfo  {journal} {J.~Mol.~Graph.}\
  }\textbf {\bibinfo {volume} {14}},\ \bibinfo {pages} {33} (\bibinfo {year}
  {1996})}\BibitemShut {NoStop}%
\end{thebibliography}

\end{document}